\documentclass{appolb}
\usepackage{graphicx}

\def\gsim{\mathrel{\rlap{\lower4pt\hbox{\hskip1pt$\sim$}}
 \raise1pt\hbox{$>$}}}

 \newcommand\beq{\begin{equation}}
 
 \newcommand\eeq{\end{equation}}
 \newcommand\beqn{\begin{eqnarray}}
 \newcommand\eeqn{\end{eqnarray}}

\def\fm{\,\mbox{fm}}
\def\GeV{\,\mbox{GeV}}
\def\TeV{\,\mbox{TeV}}
\def\lsim{\mathrel{\rlap{\lower4pt\hbox{\hskip1pt$\sim$}}
    \raise1pt\hbox{$<$}}}         
\def\gsim{\mathrel{\rlap{\lower4pt\hbox{\hskip1pt$\sim$}}
    \raise1pt\hbox{$>$}}}         

\def\fm{\,\mbox{fm}}
\def\GeV{\,\mbox{GeV}}

\begin{document}
\title{The Low theorem for diffractive bremsstrahlung and the soft photon puzzle
\thanks{Presented at ``Diffraction and Low-$x$ 2022'', Corigliano Calabro (Italy), September 24-30, 2022.}%
}
\author{B.~Z.~Kopeliovich, I.~K.~Potashnikova, Ivan~Schmidt
\address{Departamento de F\'{\i}sica,
Universidad T\'ecnica Federico Santa Mar\'{\i}a,
Avenida Espa\~na 1680, Valpara\'iso, Chile}
}
\maketitle
\begin{abstract}
The anomalous excess of small-$k_T$ photons radiated along with multi-hadron production, is challenging the physics community over four decades, but no solution has been proposed so far. We argue that the problem is rooted in the comparison with incorrect calculations, based on the so-called ``bremsstrahlung model'' (BM). It is believed to be an extension of the Low theorem from the $2\to 2+\gamma$ process to radiative multi-particle production $2\to n+\gamma$, where either initial, or final charged hadrons participate in radiation. We demonstrate that this breaks down unitarity of the $S$-matrix, thus contradicting the optical theorem.
\end{abstract}
  
\section{Low theorem revisited}

\begin{figure}[!htbp]
\begin{center}
\includegraphics[width=8cm]{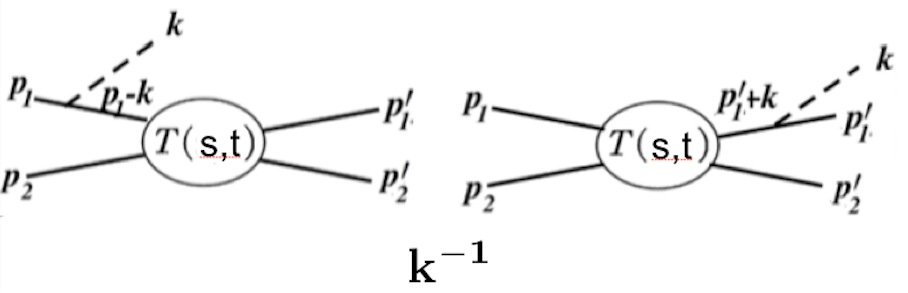}\\
{\includegraphics[width=3.5cm]{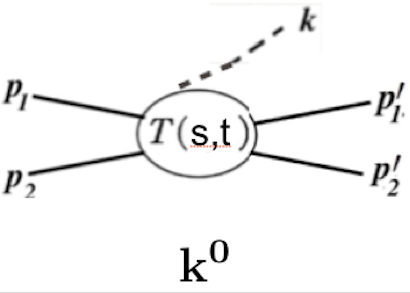}}
\caption{Graphical representation of the process (\ref{2-3}) with external (upper graphs) and intrinsic (bottom) radiation.}
\end{center}
\label{fig:2-1graphs}
\end{figure}

The process considered in the Low's paper \cite{low} is  hadronic $2\to2$ scattering, supplemented with soft photon radiation,
\beq 
h_1+h_2\to h_1^\prime+h_2^\prime+\gamma,
\label{2-3}
\eeq
with corresponding 4-momenta, $p_1,\ p_2,\ p_1^\prime,\ p_2^\prime$ and $k$.
The process is depicted in Fig.~\ref{fig:2-1graphs},
where the short-range hadronic interaction is illustrated by the blob $T(s,t)$, with $s$ and $t$ which are center-of-mass energy and 4-momentum transfer squared, respectively.

\section{Time scales}
The process (\ref{2-3}) is characterized by two time scales: 

{\bf(i)} 
One is the  coherence length of photon radiation \cite{lp} (defined by Low \cite{low} as ``the distance a particle can move with energy imbalance''), 
\beq 
l_c^\gamma = \frac{2E_1x_1(1-x_1)}{k_T^2+x_1m_h^2}.
\label{lc}
\eeq
Here $k_T$ is the photon transverse momentum; $E_1$ is the energy of $h_1$ in the rest frame of $h_2$.  The fractional light-cone momentum $x_1$ of the hadron $h_1$, carried by the photon,
\beq
x_1=\frac{k^{\gamma}_+}{p^{h_1}_+}
\label{x1}
\eeq
is invariant relative to longitudinal Lorentz boosts. 
However the hadron and photon energies, $E_1$ and  $E_\gamma$ respectively, are not invariant, so they should be taken  in Eq.~(\ref{lc}) within the same reference frame, as well as $l_c^\gamma$;

{\bf(ii)} 
At the same time, the range of strong interactions, responsible for the scattering in (\ref{2-3}) is assumed to be short in comparison with $l_c^\gamma$. Typical range of strong interaction in the target rest frame is given by the inverse pion mass $l_h\sim 1/m_\pi\sim 1\fm$.
In the same frame 
\beq
l_c^\gamma =\frac{2\omega}{k_T^2+2\omega m_h^3/s},
\label{rest-frame}
\eeq
At high energies $l_c^\gamma$ depends on photon rapidity and usually is pretty long, e.g. at 
the mid-rapidity, $l_c^\gamma=\sqrt{s}/(2k_T m_h)$. Even at large $k_T\sim 1\GeV$ and energy of the LHC range, say $\sqrt{s}=8\TeV$, $l_c^\gamma\approx 800\fm$.

Thus, we arrive at the main condition of the Low theorem,
\beq
l_c^\gamma \gg l_h.
\label{two-scales}
\eeq

Such a significant distinction between the radiation and interaction length scales 
allows to classify radiation as external and internal, respectively.

\section{"External" vs "Internal" radiation}

External bremsstrahlung is related to the radiation by the incoming and outgoing hadrons $h_1$ and $h_1^\prime$, respectively (upper graphs in Fig.~\ref{fig:2-1graphs}), while radiation from the interaction blob (bottom graph) we classify as internal. The  amplitude of the process (\ref{2-3}) is proportional to $e^\mu M_\mu$, where $e^\mu$ is the photon polarization vector, and $M_\mu=M_\mu^{ext}+M_\mu^{int}$. The first term is  infra-red divergent and reads \cite{low},
\beq
M_\mu^{ext}= \left(\frac{p_{1\mu}^\prime}{p_1^\prime k}-
\frac{p_{1\mu}}{p_1 k}\right)\,T(s,t).
\label{amplitude}
\eeq
Like in \cite{low}, for the sake of simplicity, we assume that $h_1$ is spin-less and $h_2$ is electrically neutral. Moreover, we focus on the high-energy behavior of the process (\ref{2-3}), where the energy dependence of $T(s,t)$ is very weak and can be neglected. Indeed, the extra term with the derivative $dT/ds$ included in \cite{low}, steeply vanishes with $s$. If $T(s)\propto s^\epsilon$, $dT/ds \propto s^{\epsilon-1}$ and can be safely ignored. 
So expression  (\ref{amplitude}) is full and contains no other terms.

The external radiation term is enhanced at small $k$ due to the divergent pre-factor in parentheses in Eq.~(\ref{amplitude}). It comes from the propagators of the hadrons $h_1^\prime$ and $h_1$ before and after interaction respectively.

The second term $M_\mu^{int}$ is related to $M_\mu^{ext}$ by charge conservation  $k^\mu M_\mu=0$. So it is finite at $k\to 0$, i.e. suppressed in comparison with external radiation. This is the key observation of the Low's paper.

The Low theorem can be also treated as a formal proof of the Landau-Pomeranchuk principle \cite{lp}, which states that any variation of the electric current within a finite distance does not affect the spectrum of radiation at much longer coherence length, Eq.~(\ref{lc}). Important is to keep the incoming ($l<-l_c^\gamma$) and outgoing ($l>l_c^\gamma$) currents unaffected by the current variations on shorter length scales. This means that only extrinsic radiation from initial and final hadrons $h_1$, $h_1^\prime$ matters.

\section{Photon production in inelastic collisions}

\begin{figure}[!htbp]
\begin{center}
{\includegraphics[width=4cm]{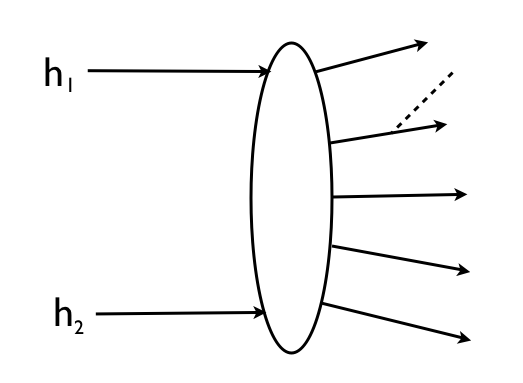}}
\caption{\label{fig:inelastic} Multi-hadron production and photon radiation in inelastic collision, as is assumed within the BM.}
\end{center}
\end{figure}

The so-called bremsstrahlung model (BM) \cite{BM}, pretending to extend the Low theorem, proven only for radiation in elastic scattering, to inelastic collisions with multi-particle production as is depicted in Fig.~\ref{fig:inelastic}, strictly contradicts data on radiative multi-particle production (see \cite{klaus}).
Like in Fig.~\ref{fig:2-1graphs}, the photon is assumed to be radiated by participating charge particles, either the incoming, or outgoing. Similar to Eq.~(\ref{amplitude}), each radiation acquires an infra-red divergent Feynman propagator. 
Such an ``extension'' of the Low theorem \cite{BM} is unjustified, and strictly contradicts the optical theorem, as demonstrated below.

\section{Unitarity relation}

\begin{figure}[!htbp]
\begin{center}
{\includegraphics[width=5cm]{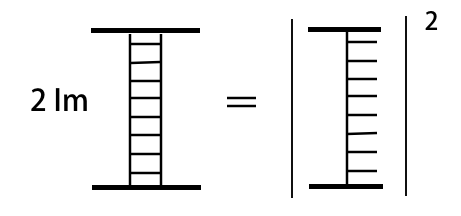}}
\caption{\label{fig:unitarity} Unitarity relation between elastic and inelastic collisions.}
\end{center}
\end{figure}

The inelastic process of multi-particle production, depicted in Fig.~\ref{fig:inelastic}, is connected to the imaginary part of the forward elastic amplitude $f^{h_1h_2}_{el}$, like is illustrated in Fig.~\ref{fig:unitarity}, due to unitarity of the $S$-matrix, $S^\dagger S=I$.
The  amplitude $f^{h_1h_2}_{el}$ is nearly imaginary at high energies, and the small elastic cross section in the r.h.s. is neglected.

The essence of the unitarity relation is the connection between the inelastic cross section of production of a final state  (including production and propagation, confinement effects, mutual interaction, etc.)  with the forward elastic amplitude. Due to unitarity, it makes no difference how to cut the elastic amplitude, i.e. at which stage of time development of the final state  the cross section is calculated. E.g. it can be done at the early stage of multi-gluon radiation, or later, when gluons fragment to quark-antiquark pairs, or at the final stage of hadronization with multiple production of hadrons. Any of these cuts
satisfy the unitarity relation, due to conservation of the total probability summed over all final states.

However, photon radiation at some later stage of final state formation, like is illustrated in Fig.~\ref{fig:inelastic}, breaks down the $S$-matrix unitarity. Indeed, production of the states before, or after photon radiation in  Fig.~\ref{fig:inelastic}, have quite different probabilities.

Moreover, one can make a unitarity cut through the radiated gluons, at the early stage, when no charged hadrons is created, and no photons can be radiated, unless it has happened at the initial state.

\section{Why Feynman rules cannot be applied to the graph in Fig.~\ref{fig:inelastic}}

Feynman propagator employed for the virtual hadrons before or after interaction in Figs.~\ref{fig:2-1graphs} and \ref{fig:inelastic}, displayed in Eq.~(\ref{amplitude}), is in fact the coherence length of radiation, Eq.~(\ref{lc}). Indeed, according to the definition (\ref{lc}), we have
\beq
\frac{1}{p^2-m_h^2}= \frac{x_1(1-x_1)}{k_T^2+x_1m_h^2}=\frac{l^\gamma_c}{2E_1}.
\label{denominator}
\eeq
This expression remains correct in the infra-red regime of vanishing photon energy $\omega\to0$, considered by Low, because $x_1=\omega/E_1$ vanishes as well. In this limit $k_T=\omega$,
so (\ref{denominator}) is $\propto 1/\omega$.

\begin{figure}[!htbp]
\begin{center}
{\includegraphics[width=4cm]{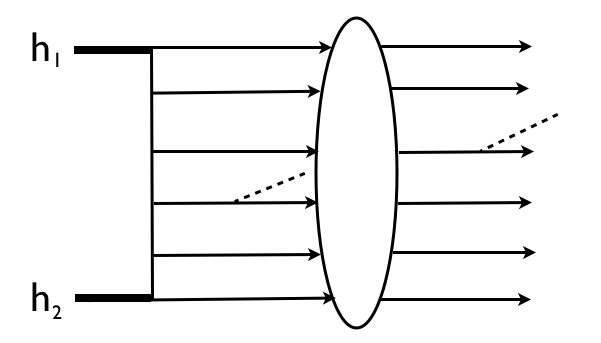}}
\caption{\label{fig:partons}Space-time pattern of particle production at high energies}
\end{center}
\end{figure}

Important is that the radiation length (\ref{lc}), i.e. the divergent Feynman propagator (\ref{denominator}),
becomes infinitely long in the infra-red limit. This demonstrates a blunder of the BM, presented in Fig~\ref{fig:inelastic}, which assumes that the outgoing particles appear momentarily from the interaction blob of finite size.
This picture contradicts the general principle of Landau-Pomeranchuk and the observation by Low \cite{low} about long coherence time of production of light particles. Within the contemporary phenomenology, like Fock-state expansion, parton model, multi-gluon radiation, etc., hadrons are composite particles, whose fluctuations have long life-time at high energy. So the constituent particles (partons) appear as fluctuations long before the interaction, which brings them to mass shell. Correspondingly, photons can be radiated either by the incoming, or outgoing charged parton, as is illustrated in Fig.~\ref{fig:partons}.
An approach, based on the quark-gluon string model, supplemented with the color-dipole model for photon radiation, was developed in \cite{michal} and led to a good description of available data.

It is also worth commenting on the choice of reference frame. In the inelastic process depicted in Fig.~\ref{fig:inelastic} there is only one natural axis, the momentum direction of the colliding hadrons. We consider only longitudinal Lorentz boosts along this direction, so $k_T$ and $x_1$ are invariant under such boosts. Apparently, small $k_T$ of the photons relative to this axis does not mean smallness of $k_T$ relative the momentum of the radiating hadron. According to Eq.~(\ref{lc}), the infra-red behavior $1/\omega$ of the propagator means vanishing $x_1$ as well. However, experimentally detected photons may have a small $k_T$, but usually they are radiated with rather large $x_1$.

\section{Conclusions}
\begin{itemize}
\item
The observed enhancement of low-$k_T$ photons, in comparison with incorrect calculations, should not be treated as a puzzle.
\item
The paper by Low considered a large rapidity gap process of diffractive excitation of a hadron, $h_1\to h_1+\gamma$, which has little to do with multiple hadron production spanning all over the rapidity interval between colliding hadrons.
\item
Unitarity between the elastic amplitude and multi-particle production cross section is independent of the stage of multi-particle formation. Photon radiation by the final state hadrons violates unitarity, because makes the relation dependent on the stage where the cut is made.

\end{itemize}
{\bf Acknowledgements:}
This work was supported in part by grant ANID PIA/APOYO AFB220004.

{}

\end{document}